\documentclass[graybox]{svmult}
\usepackage{type1cm}
\usepackage{makeidx}
\usepackage{graphicx}
\usepackage{multicol}
\usepackage[bottom]{footmisc}
\usepackage{newtxtext}
\usepackage{newtxmath}

\makeindex
\begin{document}
%\newpage
%\pagenumbering{roman}
%\addcontentsline{toc}{chapter}{Synopsis}
%\input{mytexfiles/synps}
%\newpage
%\pagenumbering{arabic}
%__________________________________________________________________
\title*{New signatures of phase transition from Statistical  Models of Nuclear multifragmentation}
\author{ G. Chaudhuri, S. Mallik, P. Das, S. Das Gupta}
\institute{G. Chaudhuri \at Variable Energy Cyclotron Centre, 1/AF Bidhan Nagar, Kolkata 700064, India\\
\email{gargi@vecc.gov.in}
\and S.Mallik \at Variable Energy Cyclotron Centre, 1/AF Bidhan Nagar, Kolkata 700064, India
\and P.Das \at Variable Energy Cyclotron Centre, 1/AF Bidhan Nagar, Kolkata 700064, India
\and S. Das Gupta \at Physics Department, McGill University, Montr{\'e}al, Canada H3A 2T8}

\maketitle

\abstract{: The study of liquid-gas phase transition in heavy ion collisions has generated a lot of interest amongst the nuclear physicists in the recent years.  In heavy ion collisions, there is no direct way of measuring  the state variables like entropy, pressure, energy and hence unambiguous characterization of phase transition  becomes difficult. This work proposes new signatures of phase transition that can be extracted from the observables which are easily accessible in experiments. It is observed that the temperature dependence of the  first order derivative of the order parameters in nuclear liquid gas phase transition exhibit similar behavior as that of the variation of specific heat at constant volume $C_v$ which is  an established signature of first order phase transition. This motivates us to propose these derivatives as confirmatory signals of liquid-gas phase transition.  The measurement of these signals in easily feasible in  most experiments as compared to the other signatures like specific heat, caloric curve or bimodality. Total multiplicity, size of largest cluster are some of the order parameters which have been studied. Statistical Models based on canonical ensemble and lattice gas model has been used for the study.  This temperature where the peak appears is designated to be the transition temperature and the effect of certain parameters on this has also been examined. The multiplicity derivative signature proposed in this work has been further confirmed by other theoretical models as well as in experimental study.}

%\maketitle
\section{Introduction}
The phenomenon of liquid-gas phase transition occurring in heavy ion collisions at intermediate energies is a subject of contemporary interest \cite{Gross, Siemens,Bondorf1,Dasgupta_Phase_transition,Borderie2,Borderie,Gross_phase_transition,Chomaz}.
The nature of nucleon-nucleon strong interaction potential, which is an attractive one with a repulsive core is very similar to the Van der Waals potential\cite{Dasgupta_Phase_transition} except for the magnitude. This type of interaction explains the phenomenon of phase transition in ordinary liquid and hence similar observation is very much expected in nuclear systems also. In ordinary liquids, when it is heated , the temperature rises till the boiling point is reached after which it remains constant till the whole amount of liquid is converted to gas. Similarly in order to observe phase transition in nuclear system, one has to pump energy to the system and only possible way is by means of nucleus-nucleus collision. In very high energy heavy ion collision at high density and temperature, hadronic matter transforms to the Quark Gluon Plasma (QGP) phase.   At the intermediate energy regime, nuclear multifragmentation is the dominant mechanism which can be related to a liquid-gas kind of transition at sub-saturation nuclear density . Theoretical models of multifragmentation predict the existence of phase transition in infinite nuclear matter. Experimental signatures also indicate change of state and this can be interpreted as finite size counterpart of the 1st order phase transition in nuclear matter. Different signatures of this transition have been studied extensively both theoretically as well as experimentally \cite{Dasgupta_Phase_transition,Borderie2,Gross_phase_transition,Chomaz,Mallik10}. The variation of excitation energy and specific heat with temperature are two well studied signatures theoretically in order to detect the first order phase transition\cite{Bondorf_NPA,Pochodzalla,Das2}.
%____________________________________________________________________
\begin{figure}[h]
\begin{center}
\includegraphics[scale=0.90]{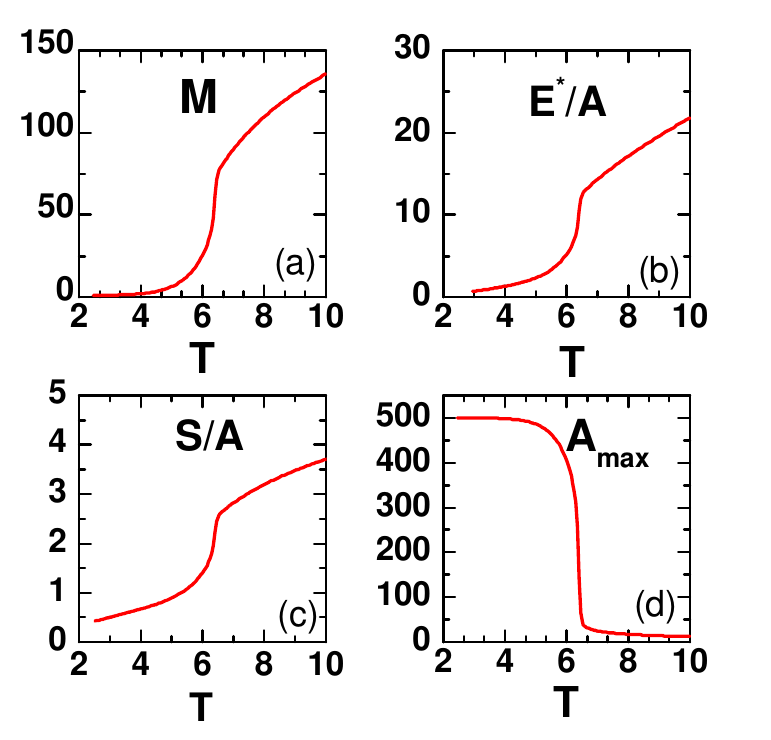}
\caption{Variation of (a)Total  Multiplicity M (b)excitation energy $E^*/A$ (MeV/A), (c) entropy per nucleon S/A (d) average size of the largest cluster $A_{max}$ with temperature T (MeV) for the fragments produced in the fragmentation of an ideal one-component system of size A=500.}
\label{fig1}
\end{center}
\end{figure}
%____________________________________________________________________

Phase transition is usually characterized by the specific behaviour of state variables like pressure, density, energy, entropy etc \cite{Reif,Pathria}. The order of phase transition, according to Ehrenfest, is determined by the lowest order derivative of free energy that shows a discontinuity.   In heavy ion collisions there is no direct way of accessing these state variables and hence unambiguous detection of phase transition  becomes difficult. The present work  is motivated by this limitation and aims at looking for  signatures of phase transition  that can be extracted from the observables which are easily accessible in experiments. Ideally phase transition exists in the thermodynamic limit and for a first order one, entropy should have a finite discontinuity and specific heat a divergence at the phase transition temperature.  In finite nuclei the discontinuity or divergence is replaced by sudden jump or maxima. The variation of total multiplicity or size of the largest cluster ($Z_{max}$)  with temperature is very much similar to that of entropy or excitation energy(caloric curve) with temperature and this can be seen from Fig. \ref{fig1} Hence  the first order derivative of these observables with temperature is expected to behave in a similar way as those of entropy or energy. This observation led to the investigation of the nature of the derivatives of these multifragmentation observables which can be easily measured in experiments. Encouraging results have been obtained from this study and it has been observed that first  order derivative of the order parameters related to the total multiplicity,  largest cluster size (produced in heavy ion collisions) exhibit similar behavior as that of the variation of specific heat at constant volume $C_v$ which is  an established signature of first order phase transition. This motivates us to propose these derivatives of total multiplicity, largest cluster size\cite{Gulminelli1,Krishnamachari,Pleimling,Chaudhuri_largest_cluster}  as confirmatory signals of liquid-gas phase transition. Another observable we have proposed here is related to the difference (normalized)  between the sizes of the first and second largest clusters which also serve as order parameter for phase transition in nuclear fragmentation and has been studied experimentally \cite{lefevre1,lefebre2} too.
 The derivatives of all these peak at the same temperature as specific heat and hence can confirm the phase transition in the fragmentation process.  The measurement of these signals in easily feasible in  most experiments as compared to the other signatures like specific heat, caloric curve or bimodality. This temperature where the peak appears is designated to be the transition temperature and the effect of certain parameters on this has also been examined.\\
 \indent
We have used mainly used statistical model based on canonical ensemble which is better known by the Canonical Thermodynamical Model (CTM)\cite{Das} in order to study the fragmentation of nuclei. In such models of nuclear disassembly it is assumed that because of multiple nucleon-nucleon collisions a statistical
equilibrium is reached and disintegration pattern is solely decided by the statistical weights in the available phase space. The temperature rises and  the
system expands from normal density and composites are formed on the
way to disassembly as a result of density fluctuation. As the system reaches between three to six times
the normal volume, the interactions between composites become
unimportant (except for the long range Coulomb interaction) and one
can do a statistical equilibrium calculation to obtain the yields of
composites at a volume called the freeze-out volume. This model can
be implemented in different statistical ensembles(microcanonical,
canonical and grand canonical \cite{Gross,Bondorf1,Das}. In our
calculation, the partitioning into available channels is solved in
the canonical ensemble where the number of particles in the nuclear
system is finite (as it would be in experiments).The study is done for different nuclear sizes, freeze out volumes and temperatures. Since Coloumb interaction is long range and suppresses the signatures of phase transition, hence we have switched off the Coloumb force in some part of our study in order to have better idea of the signatures. In such cases we have considered symmetric nuclear matter and no distinction is made between neutron and proton. In  addition to CTM we have also used the Lattice Gas Model \cite{Mallik20} recently developed in our group in order to study the multiplicity derivative signal. This model uses geometry similar to the percolation model but is much more elaborate with insertion of an Hamiltonian. Both the thermodynamic and the lattice gas model confirmed the multiplicity derivative as signature of first order phase transition in nuclear multifragmentation.

We have given brief description of the models used in our calculation in the next section. After that the results displaying the new signatures are proposed in details. The last section gives the summary of our work.
\section{Brief description of Models}
\subsection{The canonical thermodynamical model}
In this section we describe briefly the canonical thermodynamical model which is briefly designated as CTM. We assume that a system with
 $A_0$ nucleons and $Z_0$
protons at temperature $T$, has expanded to a higher than normal volume
and the partitioning into different composites can be calculated according
to the rules of equilibrium statistical mechanics.  In a canonical model, the partitioning
is done such that all partitions have the correct $A_0, Z_0$ (equivalently
$N_0, Z_0$).  Details of the implementation of the canonical model
can be found elsewhere \cite{Das}; here we give the essentials
necessary to follow the present work.

The canonical partition function is given by
\begin{eqnarray}
Q_{N_0,Z_0}=\sum\prod \frac{\omega_{I,J}^{n_{I,J}}}{n_{I,J}!}
\end{eqnarray}
Here the sum is over all possible channels of break-up (the number of such
channels is enormous) which satisfy $N_0=\sum I\times n_{I,J}$
and $Z_0=\sum J\times n_{I,J}$; $\omega_{I,J}$
is the partition function of one composite with
neutron number $I$ and proton number $J$ respectively and $n_{I,J}$ is
the number of this composite in the given channel.
The one-body partition
function $\omega_{I,J}$ is a product of two parts: one arising from
the translational motion of the composite and another from the
intrinsic partition function of the composite:
\begin{eqnarray}
\omega_{I,J}=\frac{V_f}{h^3}(2\pi mT)^{3/2}A^{3/2}\times z_{I,J}(int)
\end{eqnarray}
Here $A=I+J$ is the mass number of the composite and
$V_f$ is the volume available for translational motion; $V_f$ will
be less than $V$, the volume to which the system has expanded at
break up. We use $V_f = V - V_0$ , where $V_0$ is the normal volume of
nucleus with $Z_0$ protons and $N_0$ neutrons.  In this calculation we
have used a fairly typical value $V=6V_0$.

The probability of a given channel $P(\vec n_{I,J})\equiv P(n_{0,1},
n_{1,0},n_{1,1}......n_{I,J}.......)$ is given by
\begin{eqnarray}
P(\vec n_{I,J})=\frac{1}{Q_{N_0,Z_0}}\prod\frac{\omega_{I,J}^{n_{I,J}}}
{n_{I,J}!}
\end{eqnarray}
The average number of composites with $I$ neutrons and $J$ protons is
seen easily from the above equation to be
\begin{eqnarray}
\langle n_{I,J}\rangle=\omega_{I,J}\frac{Q_{N_0-I,Z_0-J}}{Q_{N_0,Z_0}}
\end{eqnarray}
The constraints $N_0=\sum I\times n_{I,J}$ and $Z_0=\sum J\times n_{I,J}$
can be used to obtain different looking but equivalent recursion relations
for partition functions%\cite{Chase}.  For example
\begin{eqnarray}
Q_{N_0,Z_0}=\frac{1}{N_0}\sum_{I,J}I\omega_{I,J}Q_{N_0-I,Z_0-J}
\end{eqnarray}
These recursion relations allow one to calculate $Q_{N_0,Z_0}$

We list now the properties of the composites used in this work.  The
proton and the neutron are fundamental building blocks
thus $z_{1,0}(int)=z_{0,1}(int)=2$
where 2 takes care of the spin degeneracy.  For
deuteron, triton, $^3$He and $^4$He we use $z_{I,J}(int)=(2s_{I,J}+1)\exp(-
\beta E_{I,J}(gr))$ where $\beta=1/T, E_{I,J}(gr)$ is the ground state energy
of the composite and $(2s_{I,J}+1)$ is the experimental spin degeneracy
of the ground state.  Excited states for these very low mass
nuclei are not included.
For mass number $A=5$ and greater we use
the liquid-drop formula.  For nuclei in isolation, this reads ($A=I+J$)
\begin{eqnarray}
z_{I,J}(int)=\exp\frac{1}{T}[W_0A-\sigma(T)A^{2/3}-\kappa\frac{J^2}{A^{1/3}}
-C_s\frac{(I-J)^2}{A}+\frac{T^2A}{\epsilon_0}]
\end{eqnarray}
The derivation of this equation is given in several places
\cite{Bondorf1,Das}
so we will not repeat the arguments here.  The expression includes the
volume energy, the temperature dependent surface energy, the Coulomb
energy and the symmetry energy.  The term $\frac{T^2A}{\epsilon_0}$
represents contribution from excited states
since the composites are at a non-zero temperature.

We also have to state which nuclei are included in computing $Q_{N_0,Z_0}$
(eq.(17)).
For $I,J$, (the neutron and the proton number)
we include a ridge along the line of stability.  The liquid-drop
formula above also gives neutron and proton drip lines and
the results shown here include all nuclei within the boundaries.

The long range Coulomb interaction between
different composites can be included in an approximation called
the Wigner-Seitz approximation.  We incorporate this following the
scheme set up in \cite{Bondorf1}.

\subsection{The evaporation code}
The statistical multifragmentation model described above calculates the
properties of the collision averaged system that can be approximated by an
equilibrium ensemble. Ideally, one would like to measure the properties of
excited primary fragments after emission in order to extract information about
the collisions and compare directly with the equilibrium predictions of the
model. However, the time scale of a nuclear reaction($10^{-20}s$) is much
shorter than the time scale for particle detection ($10^{-9}s$). Before reaching
the detectors, most fragments decay to stable isotopes in their ground states.
Thus before any model simulations can be compared to experimental data, it is
indispensable to have a model that simulates sequential decays.
A Monte Carlo technique is employed to follow all decay chains until the
resulting products are unable to undergo further decay. For the purposes of the
sequential decay calculations  the excited primary fragments generated by the
statistical model calculations are taken as the compound nucleus input to the
evaporation code. Hence, every primary fragment is decayed as a separate event.

We consider the deexcitation of a primary fragment of mass $A$, charge $Z$
and temperature $T$. The successive particle emission from  the hot primary fragments
is assumed to be the basic deexcitation mechanism. For each event of the primary
breakup simulation, the entire chain of evaporation and secondary breakup events
is Monte Carlo simulated. The standard Weisskopf evaporation
 scheme is used to take into account evaporation of nucleons, $d$, $t$,
 $He^3$ and $\alpha$. The decays of particle stable excited states via gamma
 rays were also taken into account for the sequential decay process and for the
 calculation of the final ground state yields. We have also considered fission
 as a deexcitation channel though for the nuclei of mass $<$ 100 its role will be
quite insignificant. The process of light particle emission from a compound nucleus is
governed by the emission width $\Gamma_{\nu}$ at which a
particle of type $\nu$ is emitted. The different equations for calculation of particle, gamma and fission widths is given in details in \cite{Mallik1} and we will skip them here.
 Once the emission widths are known, it is required to
establish the emission algorithm which decides  whether a particle is
 being emitted from the compound nucleus.
 This is done \cite{thesis} by first
calculating the ratio $x=\tau / \tau_{tot}$  where $\tau_{tot}=
\hbar  / \Gamma_{tot}$, $\Gamma_{tot}=\sum_{\nu}\Gamma_{\nu}$ and
$\nu = n,p,d,t,He^3,\alpha,\gamma$ or fission and then performing Monte-Carlo
sampling from a uniformly distributed set of random numbers.
 In
the case that a particle is emitted, the type of the emitted
particle is next decided by a Monte Carlo selection with the
weights $\Gamma_{\nu}/\Gamma_{tot}$ (partial widths).
 The energy of the emitted particle is then obtained by
another Monte Carlo sampling of its energy spectrum. The energy, mass and charge
of the nucleus is adjusted after each emission. This procedure is followed for
each of the primary fragment produced at a fixed temperature and then
repeated over a large ensemble and the observables are calculated from the
ensemble averages. The number and type of particles emitted and the
final decay product in each event is registered and are taken into account  properly
keeping in mind the overall charge and baryon number conservation.

\subsection{Lattice Gas Model}
The Lattice Gas Model is considerably more complicated than the percolation model\cite{Mallik20} but expositions of the model exist \cite{Chomaz, Pan,Samaddar,Chomaz} and we refer to \cite{Samaddar} for details.  Let $A=N+Z$ be the number of nucleons in the system that dissociates.  We consider $D^3$ cubic boxes where each cubic box has volume $(1.0/0.16)fm^3$. $D^3$ is larger than $A$ (they have the same value in bond percolation model).
Here $D^3/A=V_f/V_0$ where $V_0$ is the normal volume of a nucleus with $A$ nucleons and $V_f$
is the freeze-out volume where partitioning of nucleons into clusters is computed.  For nuclear forces one adopts nearest neighbor interactions.  Following normal practice, we use neutron-proton interactions $v_{np}$=-5.33 MeV and set $v_{nn}=v_{pp}$=0.0. Coulomb interaction between protons is included.  Each cube can contain 1 or 0 nucleon. There is a very large number of configurations that are possible (a configuration designates which cubes are occupied by neutrons, which by protons and which are empty; we sometimes call a configuration an event).  Each configuration has an energy.  If a temperature is specified, the occupation probability of each configuration is proportional to its energy: P$\propto$exp(-E/T).  This is achieved by Monte-Carlo sampling using Metropolis
algorithm.

%\\
%\begin{figure}[t]
%\includegraphics[width=6.5cm,keepaspectratio=true]{Paper20_Fig4.eps}
%\caption{(Color online) Variation of d$M$/d$T$ (red solid lines) and $C_v$ (green dashed lines) with temperature from lattice gas model at $D$=8 (see text) for fragmenting system %having $Z$=82 and $N$=126. To draw d$M$/d$T$ and $C_v$ in the same scale, $C_v$ is normalised by a factor of 1$/$10; d$M$/d$T$ is unit of MeV$^{-1}$.}
%\end{figure}
%\indent
Calculation of clusters need further work.  Once an event is chosen we ascribe to each nucleon a momentum.
Momentum of each nucleon is picked by Monte-Carlo sampling of a Maxwell-Boltzmann distribution for the prescribed
temperature T.  Two neighboring nucleons are part of the same cluster if $\vec{P}_r^2/2\mu+\epsilon<0$
where $\epsilon$
is $v_{np}$ or $v_{nn}$ or $v_{pp}$.  Here $\vec{P}_r$ is the relative momentum of the two nucleons and $\mu$ is
the reduced mass.  If nucleon $i$ is bound with nucleon $j$ and $j$ with $k$ then $i, j, k$ are part of the same
cluster.  At each temperature we calculate 50,000 events to obtain average energy
$<E>$ and average multiplicity $n_a$
(where $a$ is the mass number of the cluster) of all clusters.  A cluster with 1 nucleon is a monomer, one with 2
nucleons is a dimer and so on.  The total multiplicity is $M=\sum n_a$
and $\sum an_a=A$ where $A=N+Z$ is the mass number of the dissociating system.
%____________________________________________________________________
\begin{figure}[t]
\begin{center}
\includegraphics[scale=0.70]{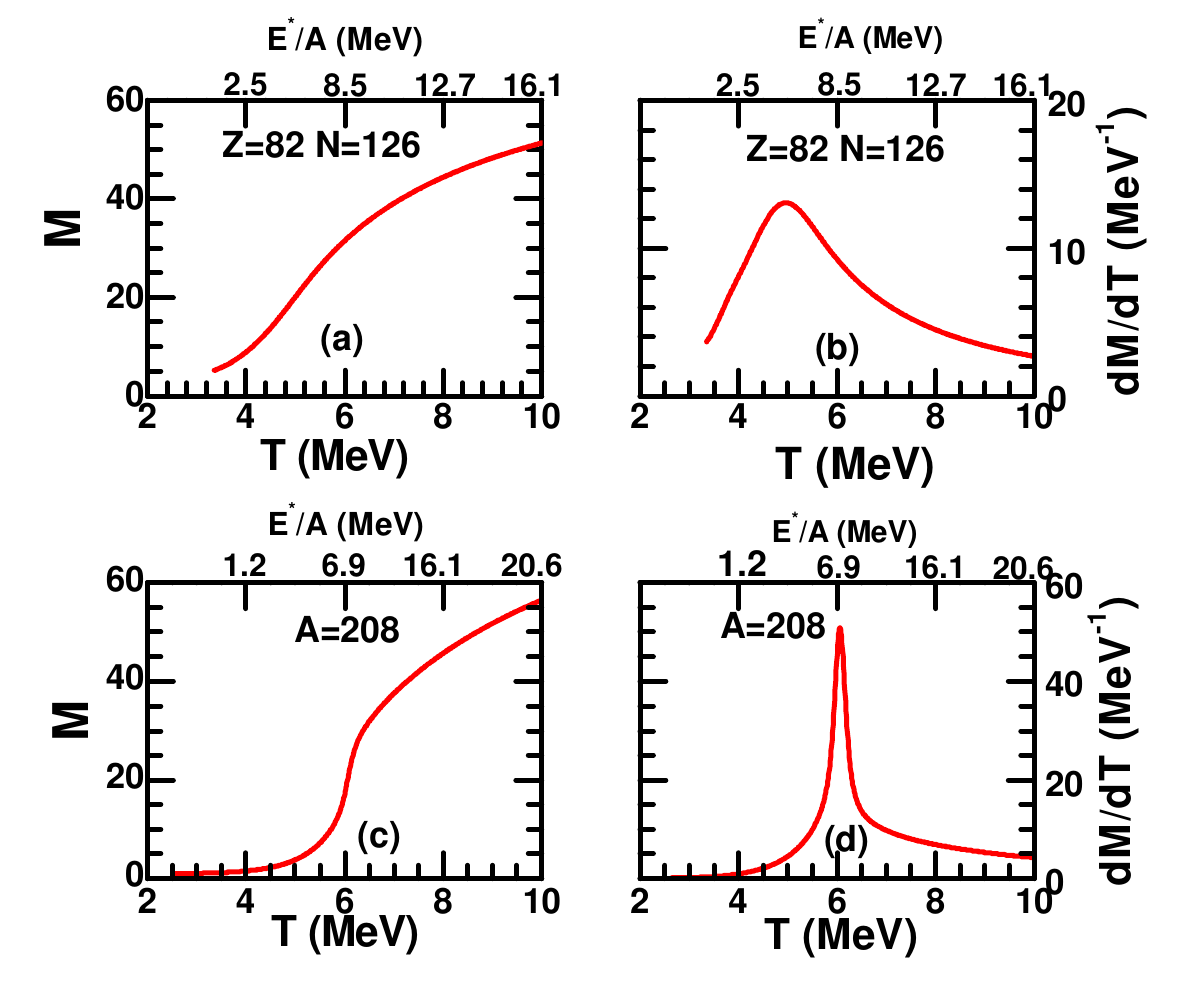}
\caption{Variation of multiplicity M ((a) and (c)) and dM/dT ((b) and (d)) with temperature (bottom x axes) and excitation per nucleon (top x axes) from the CTM calculation for fragmenting systems having Z =82 and N =126 ((a) and (b)). (c) and (d) represent the same but for a hypothetical system of one kind of particle with no Coulomb interaction but the same mass number (A =208). $E^*=E-E_0$, where $E_0$ is the ground-state energy of the dissociating system in the liquid drop model whose parameters are given in Ref. \cite{Das}}
\label{fig2}
\end{center}
\end{figure}
%____________________________________________________________________
\section{Results}
\begin{figure}[h]
\begin{center}
\includegraphics[scale=0.60]{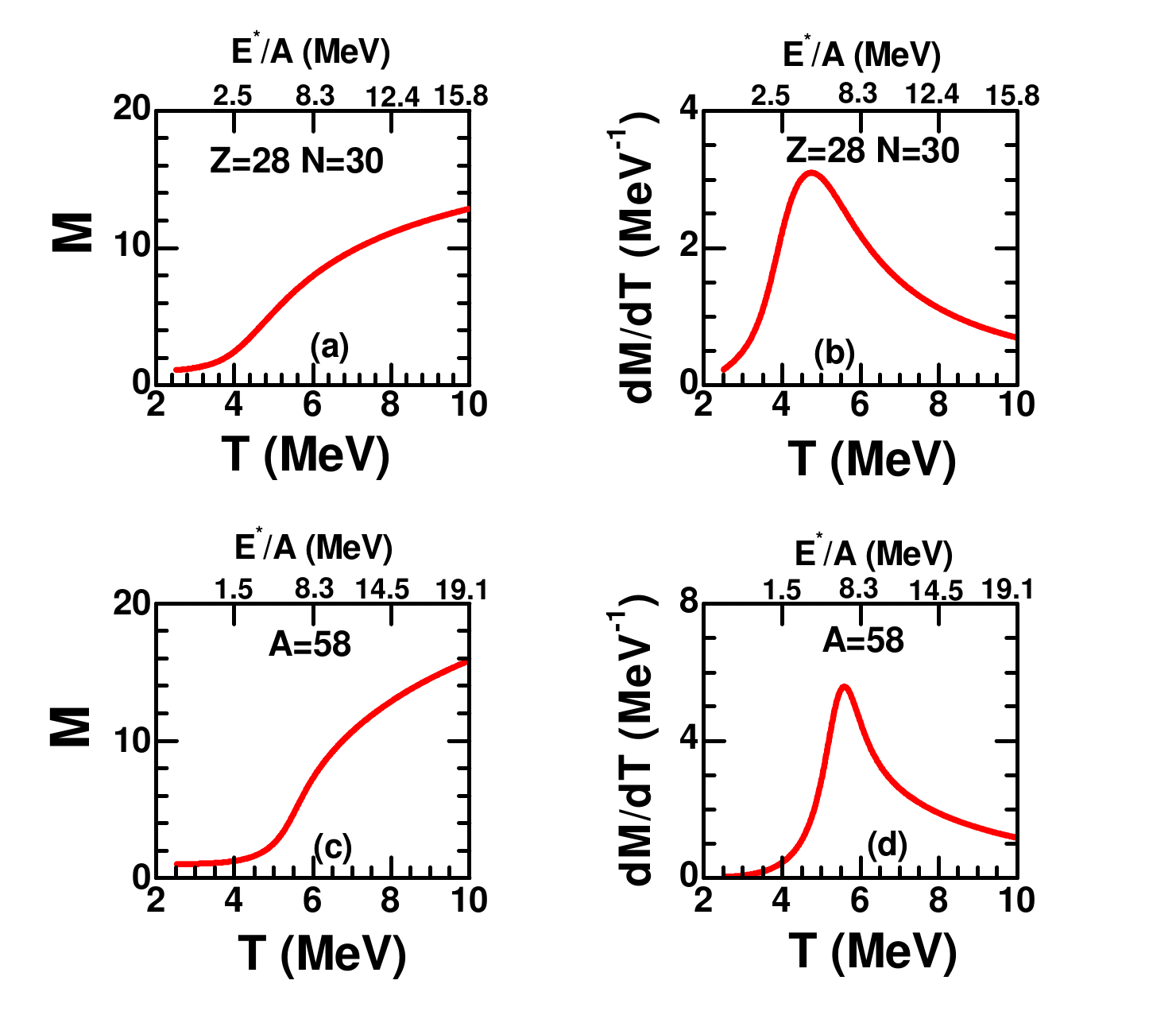}
\caption{Same as Fig.~\ref{fig2} but the fragmenting systems are Z =28 and N =30 ((a) and (b)) and A =58 ((c) and (d)).}
\label{fig3}
\end{center}
\end{figure}

The variation of total multiplicity with temperature for fragmenting systems as calculated by CTM is very similar to that of entropy. This motivated us to look for the derivative of multiplicity  which is expected to behave similarly as the derivative of entropy w.r.t temperature which is nothing but the specific heat at constant volume $C_V$. Unlike the entropy, one can measure the total multiplicity $M=\sum M_a$ ($a$ being the mass number of the composites)  with $4\pi$ detectors in the laboratory.In
CTM the derivative of $M$ with $T$ as a function of $T$ is seen to have a maximum.  Fig. \ref{fig2} (left panel) shows the total multiplicity for fragmenting system having proton number ($Z$)=82 and neutron number ($N$)=126 and and its derivative $dM/dT$ (the right panel). Results for both real nuclei and the one for one kind of particles have been displayed in order to emphasize the effects of Coulomb interaction. The rise and the peak are much sharper in absence of Coulomb interaction clearly indicating the role of the long range interaction in suppressing the signatures of phase transition. The features become less sharp as in $Z$=28 and $N$=30,  as the system size decreases (Fig. \ref{fig3}),
%\newpage
%____________________________________________________________________
%____________________________________________________________________
%____________________________________________________________________
\begin{figure}[htb]
\begin{center}
\includegraphics[scale=0.60]{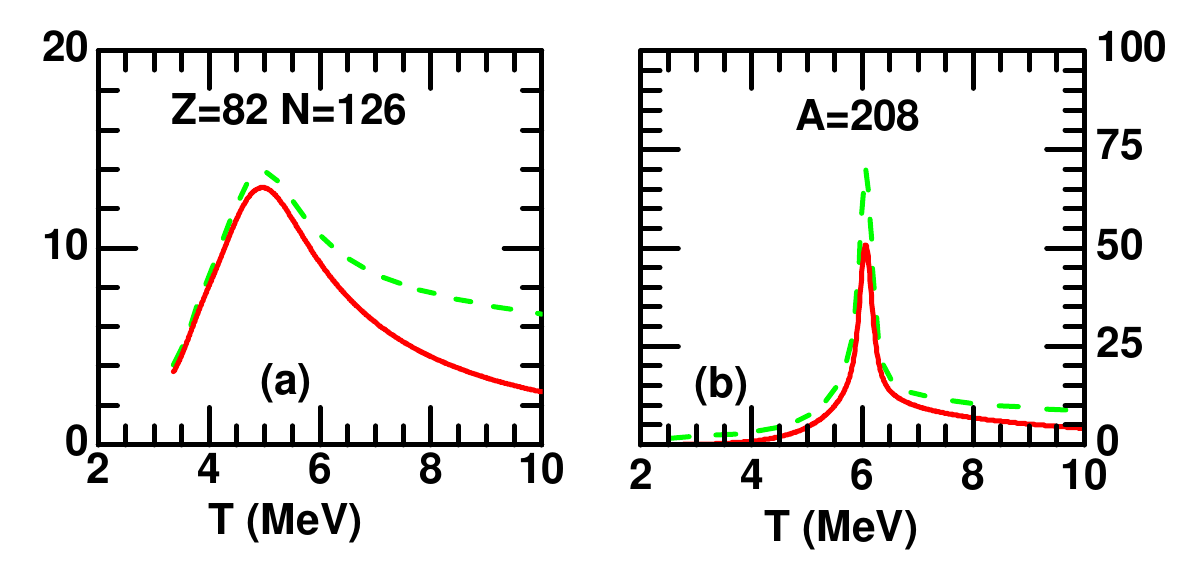}
\caption{ Variation of dM/dT (red solid lines) and $C_v$ (green dashed lines) with temperature from CTM for fragmenting systems having Z =82 and N =126 (a) and for hypothetical systems of one kind of particle with no Coulomb interaction of mass number A =208 (b). To draw dM/dT and $C_v$ in the same scale, Cv is normalized by a factor of 1/50.
 [Reprinted with permission from S Mallik, G. Chaudhuri, P. Das, S. Das Gupta, Phys. Rev. C {\bf 95}, 061601 , 2017 (R)] Copyright (2020) from the American Physical Society }
\label{fig4}
\end{center}
\end{figure}
%____________________________________________________________________

\begin{figure}[h]
\begin{center}
\includegraphics[scale=0.5]{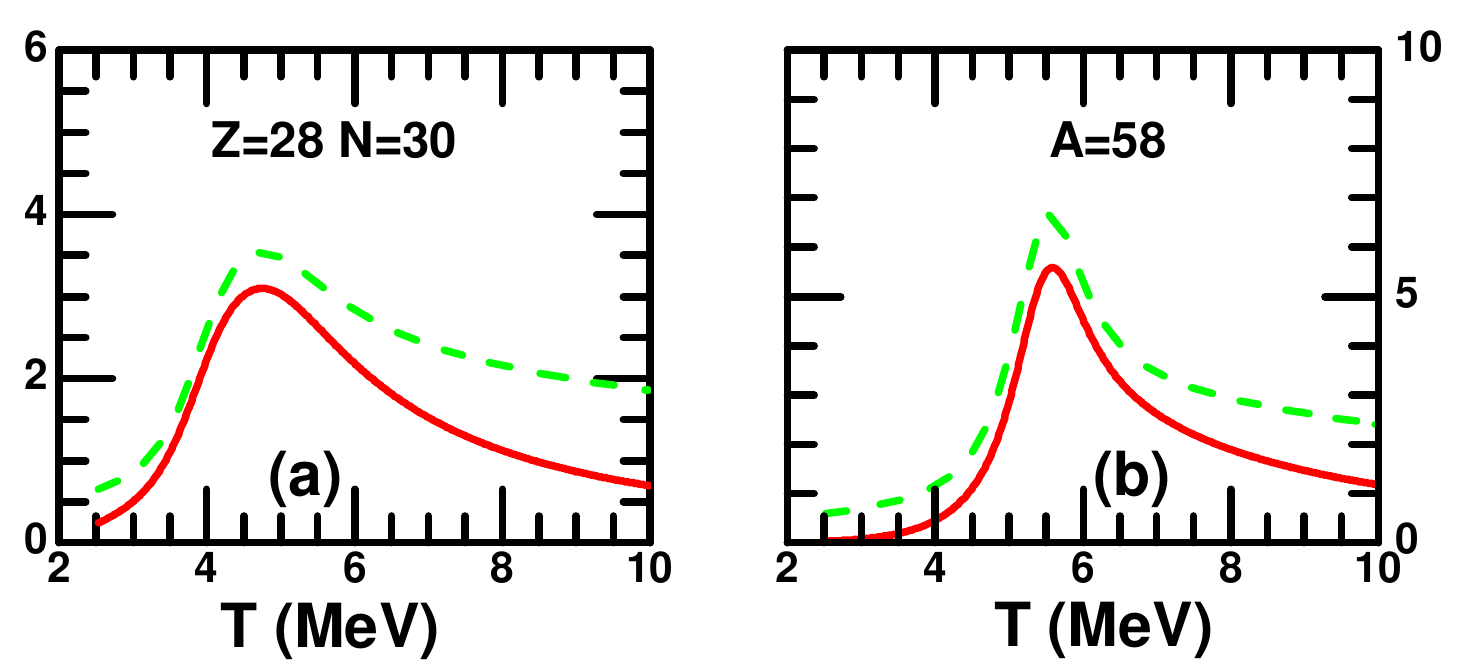}
\caption{Same as in Fig. \ref{fig4}, but the fragmenting systems are Z =28 and N =30 (a) and A =58 (b).
[Reprinted with permission from S Mallik, G. Chaudhuri, P. Das, S. Das Gupta, Phys. Rev. C {\bf 95}, 061601 , 2017 (R)] Copyright (2020) from the American Physical Society }
\label{fig5}
\end{center}
\end{figure}

\begin{figure}[hbt]
\begin{center}
\includegraphics[scale=0.80]{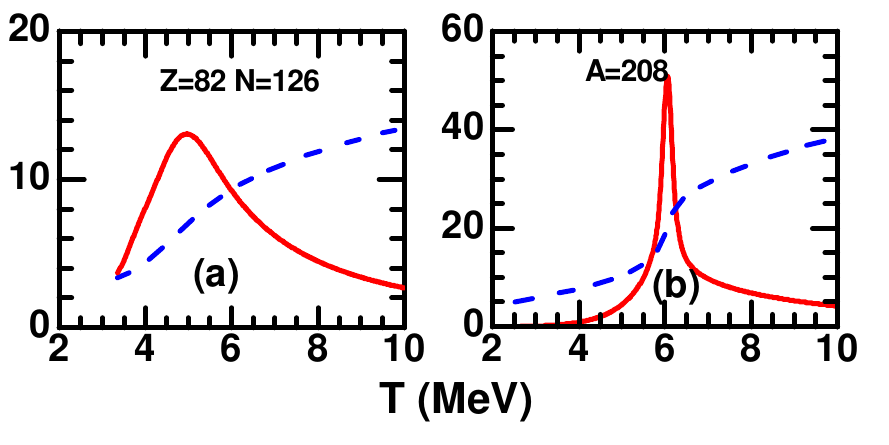}
\caption{ Variation of entropy (blue dashed lines) and dM/dT (red solid lines) with temperature from CTM for fragmenting systems having Z =82and N =126 (a) and for hypothetical system of one kind of particle with no Coulomb interaction of mass number A =208 (b). To draw S and dM/dT in the same scale, S is normalized by a factor of 1/20 for Z =82 and N =126 system and 1/50 for hypothetical system of one kind of particle.
[Reprinted with permission from S Mallik, G. Chaudhuri, P. Das, S. Das Gupta, Phys. Rev. C {\bf 95}, 061601 , 2017 (R)] Copyright (2020) from the American Physical Society }
\label{fig6}
\end{center}
\end{figure}

In the next two figures Fig. \ref{fig4} and Fig. \ref{fig5} we compare dM/dT and $C_V$ for the two same systems as used in \ref{fig2} and \ref{fig43}. We also consider the situation where the Coulomb is switched off. The peak in $dM/dT$ coincides with the maximum of  specific heat at constant volume $C_v$  as a function of temperature for all the cases. Its an established fact that specific heat at constant volume peaks at the transition temperature and this is a signature of 1st order phase transition. Hence based on our results as presented in \ref{fig4} and \ref{fig5}, we conclude that dM/dT can be a signature of phase transition and the advantage is it gives an exact value of the transition temperature where the maximum of dM/dT occurs. Next we calculate the entropy since its well known that
it shows a sharp rise near the transition temperature ., We have compared the temperature variation of dM/dT and the entropy for the fragmenting system Z=82, N=126, and also for an ideal condition, neglecting the Coulomb interaction. This is displayed in in the next plot \ref{fig6}. It is evident, that for both the cases, the entropy changes rapidly in that regime of  temperature scale where dM/dT exhibits a maximum.. The presence of Coulomb interaction in a real system, only, smears the rise of entropy and the peak in dM/dT. This behaviour, further, establishes that dM/dT is indeed a signature of phase transition.
%____________________________________________________________________
%____________________________________________________________________

\begin{figure}[h]
\begin{center}
\includegraphics[scale=0.60]{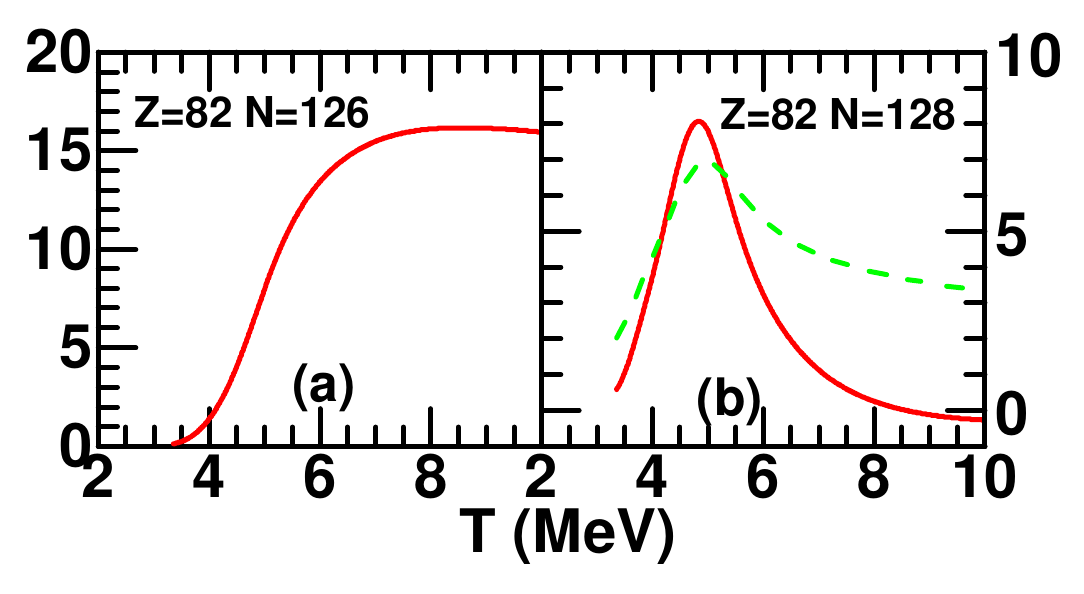}
\caption{ Variation of intermediate-mass fragment(IMF) multiplicity $M_{IMF}$ (a) and first-order derivative of IMF multiplicity $dM_{IMF}/dT$ (b) with temperature from CTM calculation for fragmenting systems having Z =82 and N =126. Variation of $C_v$ with temperature (T) is shown by green dashed line in (b). To draw d$M_{IMF}/dT$ and $C_v$ in the same scale, $C_v$ is normalized by a factor of 1/100.
[Reprinted with permission from S Mallik, G. Chaudhuri, P. Das, S. Das Gupta, Phys. Rev. C {\bf 95}, 061601 , 2017 (R)] Copyright (2020) from the American Physical Society }
\label{fig7}
\end{center}
\end{figure}

The multiplicity of the intermediate mass fragments ($M_{IMF}$) in heavy ion collisions strongly confirms the process of multifragmentation \cite{Bondorf1}.
 It is an important observable of multifragmentation, which is measured in the experiment, sometimes, instead of the total multiplicity M. Therefore,  we wanted to perform a similar test on the derivative  of $M_{IMF}$. We have plotted the variation of $M_{IMF}$ and its temperature derivative with temperature for the system Z=82, N=126 in \ref{fig7}, and compared $dM_{IMF}$/dT with $C_V$. $M_{IMF}$ and $dM_{IMF}$/dT display a similar behaviour as that of the total multiplicity and its derivative except for the fact that the peak position of its derivative do not coincide with that of $C_V$.  This is expected because the calculation of $C_V$ involves all the fragments irrespective of their mass or charge, but in $M_{IMF}$, only selected fragments are included.
 %____________________________________________________________________
%____________________________________________________________________

%____________________________________________________________________
%______________
%____________________________________________________________________
\begin{figure}[h]
\begin{center}
\includegraphics[scale=0.60]{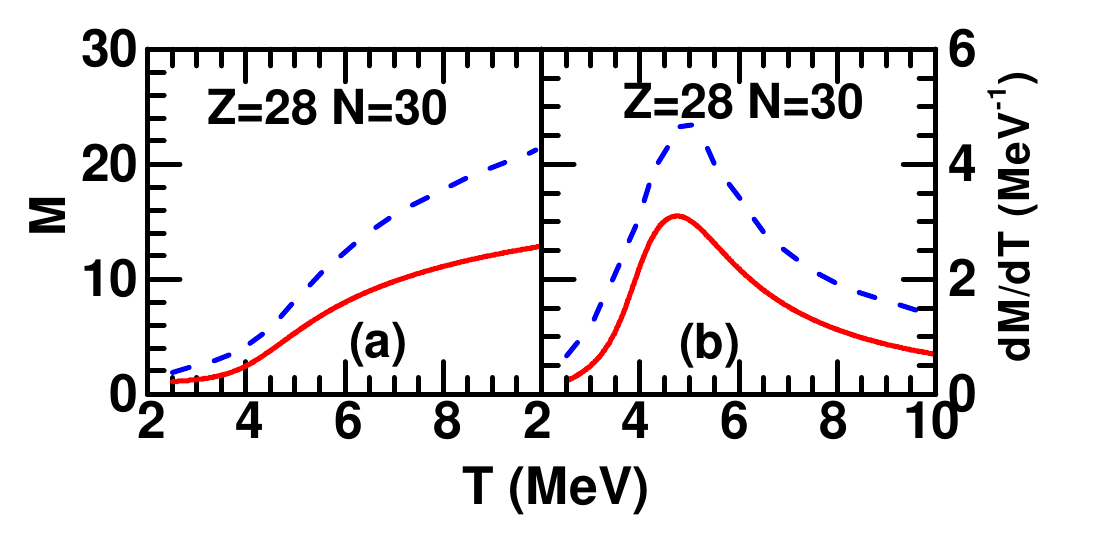}
\caption{ Effect of secondary decay on M (a) and dM/dT (b) for fragmenting systems having Z =28 and N =30. Red solid lines show the results after the multifragmentation stage (calculated from CTM), whereas blue dashed lines represent the results after secondary decay of the excited fragments.
[Reprinted with permission from S Mallik, G. Chaudhuri, P. Das, S. Das Gupta, Phys. Rev. C {\bf 95}, 061601 , 2017 (R)] Copyright (2020) from the American Physical Society }
\label{fig8}
\end{center}
\end{figure}
%____________________________________________________________________

%____________________________________________________________________
\begin{figure}[h]
\begin{center}
\includegraphics[scale=1.0]{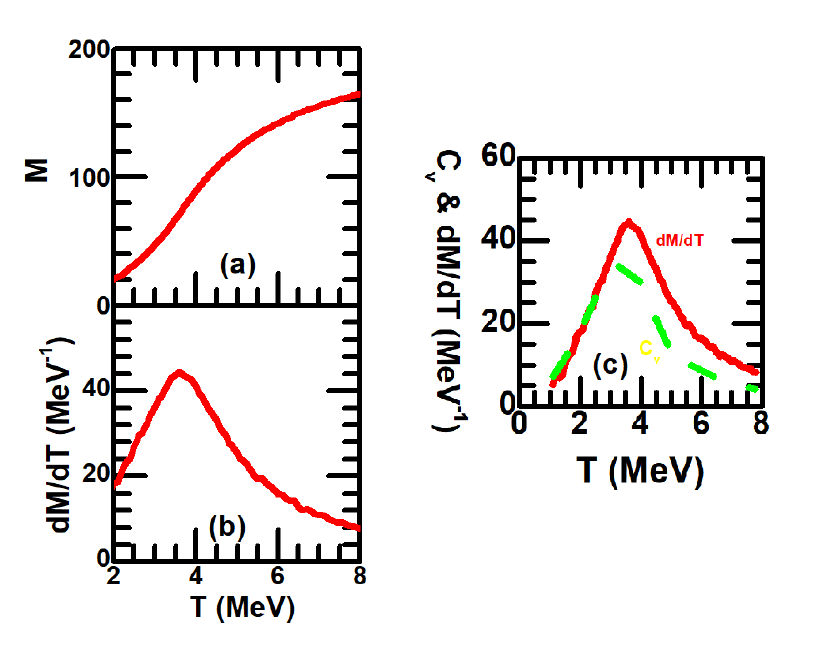}
\caption{Variation of  M  (a)and dM/dT (b) (red solid lines) and $C_v$  with temperature from lattice gas model at $D=8$ (see text) for fragmenting system having $Z$=82 and $N$=126.  (c) d$M$/d$T$ (red solid lines) and $C_v$ (green dashed lines) with T ; to draw them in the same scale, $C_v$  is normalised by a factor of 1$/$10; d$M$/d$T$ is unit of MeV$^{-1}$.
[Reprinted with permission from  S. Das Gupta,S. Mallik and G. Chaudhuri Phys. Rev. C {\bf 97}, 044605, 2018] Copyright (2020) from the American Physical Society }
\label{fig9}
\end{center}
\end{figure}
%____________________________________________________________________

Last but not the least we would like to study the effect of secondary decay on the excited fragments formed after multifragmentation. In a heavy ion collision, when a nucleus breaks up through the process of nuclear multifragmentation,  the resulting composites are called primary fragments. The primary fragments are excited in general, and lose excitation through sequential two-body decay, and thus change the total multiplicity. The final cold fragments, called secondary fragments, are detected in the laboratory.

 The fragments that we are dealing with in our study (using CTM), are primary fragments. The secondary decay may affect the total multiplicity in such a way that might change the behaviour of multiplicity discussed above. As we are interested in the experimental signature, we investigate the effect of the secondary decay in our calculation , and do the same study with the multiplicity of the secondary fragments.
 %\newpage
%____________________________________________________________________
\begin{figure}[h]
\begin{center}
\includegraphics[scale=0.70]{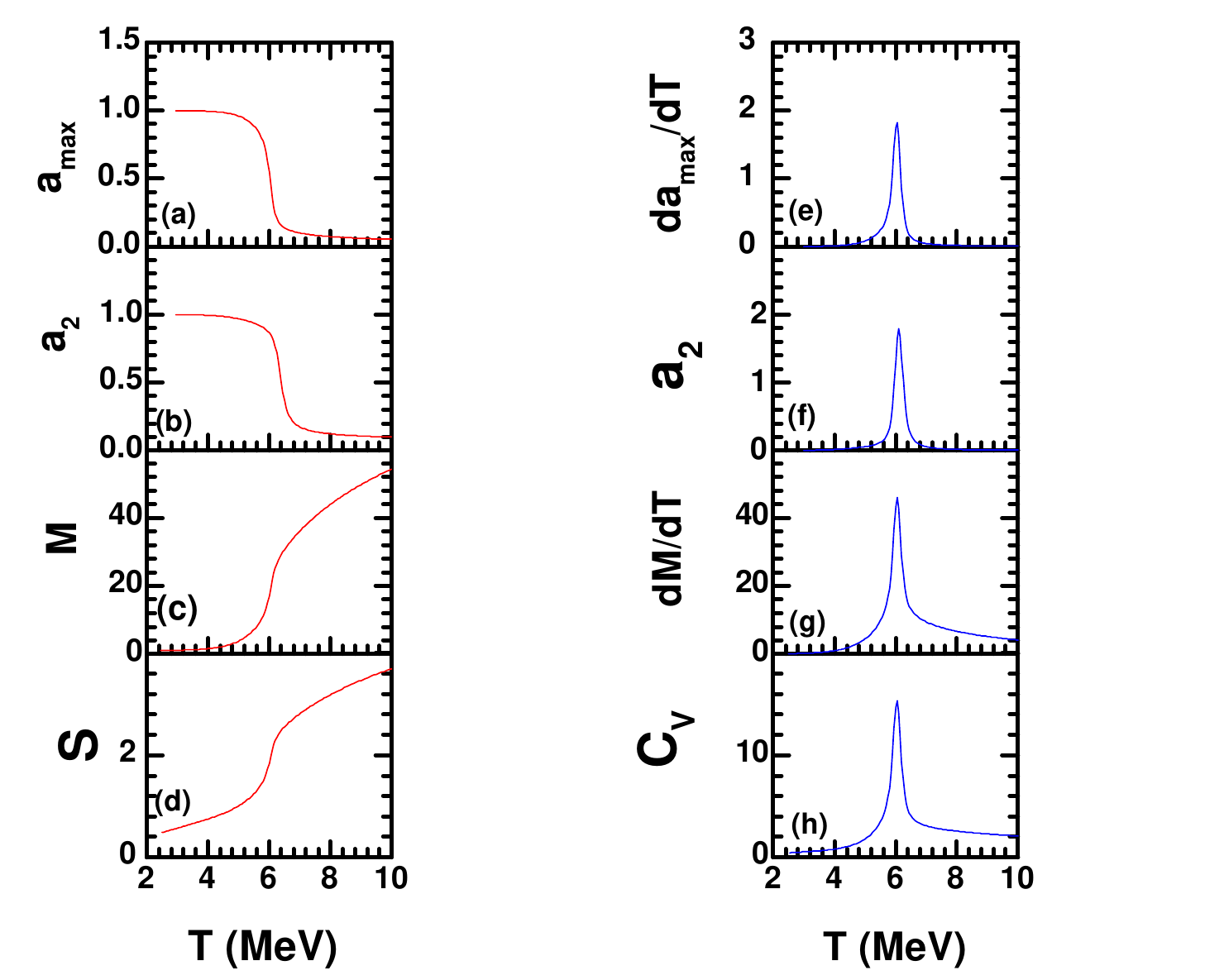}
\caption{Variation of (a) $\textbf{a}_{max}$, (b) $\textbf{a}_2$, (c) M, (d) S, (e) -$d \textbf{a}_{max}/dT$, (f) -$d \textbf{a}_2/dT$, (g) dM/dT and (h) $C_v$ with temperature for fragmenting system of mass A=200.}
\label{fig10}
\end{center}
\end{figure}
%____________________________________________________________________
%____________________________________________________________________
\begin{figure}[h]
\begin{center}
\includegraphics[scale=0.70]{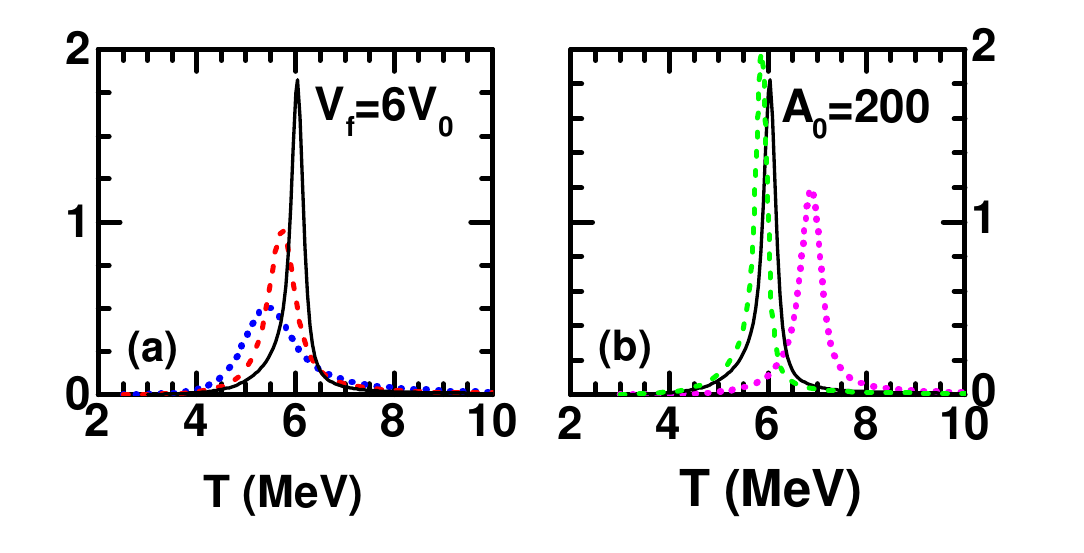}
\caption{Variation of $d\textbf{a}_{max}/dT$ with temperature (a) at constant freeze-out volume $V_f=6V_0$ but for three fragmenting system of mass 50 (blue dotted line), 100 (red dashed line) and 200 (black solid line) and (b) for same fragmenting system of mass 200 but at three constant freeze-out volumes $V_f=2 V_0$ (magenta dotted line), $V_f=6V_0$ (black solid line) anf $V_f=8V_0$ (green dashed line).}
\label{fig11}
\end{center}
\end{figure}
%____________________________________________________________________
%____________________________________________________________________
\begin{figure}[hbt]
\begin{center}
\includegraphics[scale=0.7]{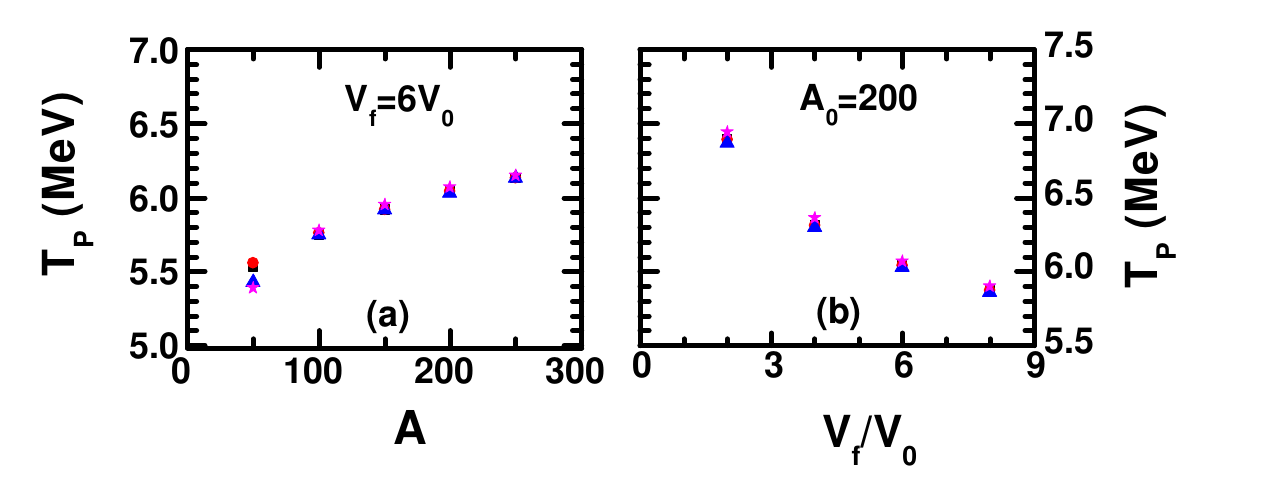}
\caption{Dependence of the peak position of -$d \textbf{a}_{max}/dT$, -$d \textbf{a}_2/dT$, dM/dT and $C_v$ on fragmenting system size (a) and freeze-out volume (b).  }
\label{fig12}
\end{center}
\end{figure}
%____________________________________________________________________

%\newpage
  We have plotted the multiplicities of the primary and the secondary fragments and their derivatives in \ref{fig8}. It is apparent that the effect of secondary decay does not alter our previous observation. Moreover, it enhances the signals, the total multiplicity changes more rapidly, and the peak in $dM/dT$ is sharper in case of the secondary fragments. Thus the maxima of multiplicity derivative can be extracted successfully through experiments with an unaltered transition temperature.

In order to further test  multiplicity derivative as a possible signature for 1st order phase transition, we have carried out the investigation using the Lattice-gas Model\cite{Mallik20}.
This is shown in \ref{fig9} We have plotted $M$ and its
derivative  against the temperature $T$ in the left panel. $M$ shows a rise and the derivative shows a peak as expected.
Plots of $dM/dT$ and $d<E>/dT$ are shown in \ref{fig9}(right panel).   $C_v$ goes through a maximum at some temperature
which is a hallmark of first order phase transition and this occurs at the same temperature where $dM/dT$
maximises.  This is remarkably similar to  results from  CTM corroborating the evidence that the appearance of a maximum in $dM/dT$ is indicative of a first
order phase transition.%\\
Our proposed signal of multiplicity derivative $dM/dT$ was tested and verified in different statistical and dynamical models like the statistical multifragmentation model (SMM) \cite{Lin1,Lin2}, Quantum Molecular Dynamics (QMD) model \cite{Liu} and Nuclear statistical Equilibrium (NSE) model\cite{Bakeer}. Our theoretical proposition of this signal got further support when it was experimentally verified recently and tested using three different reactions $^{40}Ar + ^{58}Ni$ , $^{40}Ar + ^{27}Al$  and $^{40}Ar + ^{48}Ti$ at 47 Mev/n \cite{Wada}.

 The  average size of the largest cluster $\langle A_{max} \rangle$ formed in the fragmentation of the excited nuclei acts as an order parameter for 1st order phase transition. The variable $a_2$ which is a measure of the difference between the average size of the 1st ($\langle A_{max} \rangle$) and the 2nd ($\langle A_{max-1} \rangle$) largest cluster size divided by the sum of these two (${a}_2=\frac{\langle A_{max}\rangle\,-\,\langle A_{max-1}\rangle}{\langle A_{max}\rangle\,+\,\langle A_{max-1}\rangle}$) also has similar behaviour as that of $\langle A_{max} \rangle$. So this observable which is measured in some experiments can also act as an order parameter. The analytical expressions leading to the calculation of the average size of 1st and 2nd largest cluster can be found in \cite{pdasplb}.     Now, we will concentrate on these observables in order to study their variation with temperature.
 We consider an ideal system of A=200 identical nucleons with no Coulomb force acting between them in order have better idea of these proposed signatures. Left panels of \ref{fig10}((a) to (d)) display the variations of the four variables, the normalised size of the average largest cluster $\textbf{a}_{max}$ $(\textbf{a}_{max}=\frac{\langle A_{max}\rangle}{A})$, $\textbf{a}_2$, total multiplicity M and entropy per particle (S/A) with temperature.

 $\textbf{a}_{max}$ and $\textbf{a}_2$ are almost constant and assume a value $\approx 1$ up to approximately 5 MeV, in the temperature scale. This implies that in this temperature range, the size of the largest fragment produced is almost the same as the size of the fragmenting source. Around T=6 MeV, both of them fall sharply to a very low value near zero, which indicate the entire system fragments into the light mass nuclei. After that, they remain almost unchanged. These observables, clearly, give a sharp transition near T=6 MeV and therefore behave as an order parameter of the nuclear phase transition. Now, the last two panels ((c) and (d)) in the left of Fig.10, show the variation of the total multiplicity and entropy per nucleon with temperature. $\textbf{a}_{max}$ and $\textbf{a}_2$ display similar behaviour as that of the multiplicity and the entropy; the sudden jump (or fall) of these four variables occur almost at the same temperature around 6 MeV. This similarity motivates us to investigate the behaviour of the derivatives of $\textbf{a}_{max}$ and $\textbf{a}_2$. In the right panel of f\ref{ig10}, temperature derivatives of all the four quantities are plotted as function of temperature. In the right bottom panel \ref{fig10}(h), we have plotted $C_V$, which is related to the temperature derivative of the entropy (S). The derivatives of $\textbf{a}_{max}$ and $\textbf{a}_{2}$ exhibit maxima just like total multiplicity and specific heat, and almost at the same temperature, which we call the transition temperature. This establishes these two variables as signatures of phase transition. This signature is much easier to access both theoretically and experimentally as compared to the bimodality in the probability distribution of the largest cluster. The later has been used  so far in order to detect the existence of phase transition in nuclear multifragmentation but to detect two peaks(bimodality) of equal height in a distribution at a particular temperature (or excitation energy) is far more a difficult job than to simply calculate the derivative in its size with temperature or excitation energy. We strongly believe that this new proposed signature related to the largest cluster size will definitely provide a great impetus to the study of liquid gas phase transition in heavy ion collisions.

Next, we have examined how the transition temperature  varies with the source size and the freeze-out volume. We have plotted the variation of d$\textbf{a}_{max}$/dT with T  for three different fragmenting systems of size A=50, 100, 200 at a fixed freeze-out volume $V_f=6V_0$ in \ref{fig11}(a), and the same for three freeze-out volume $V_f$=3$V_0$, $4V_0$, $8V_0$ with fixed source A=200 in \ref{fig11}(b). We see that the peaks are sharper for the more massive source and the higher freeze-out volume. The position of the peak is observed to shift to the higher temperature region for the bigger source size, and the lower temperature side for the greater freeze-out volume. This implies that the smaller system fragments more easily at a lower transition temperature as compared to its bigger counterparts. The peak also becomes sharper for bigger sources which once again proves that phase transition signals are enhanced in larger systems. For freeze-out volume, the result that we have obtained is expected, since higher freeze-out volume (lower density) will favour the disintegration of the nucleus, resulting in lower transition temperature.

At the end, we have plotted the transition temperatures as a function of system size at fixed freeze-out volume (left panel (a)), and as a function of freeze-out volume for a fixed system (right panel (b)) in f\ref{ig12}. In each panel, four different sets of transition temperatures are plotted. Those sets are obtained from the position of the maxima in $d\textbf{a}_{max}/dT$, $d\textbf{a}_{2}/dT$, dM/dT and $C_V$. The transition temperatures obtained from all the four observables give consistent results. Small differences between them  can be attributed to the finiteness of the fragmenting system.

\section{Summary} This work introduces some new signatures of nuclear liquid gas phase transition which can be measured easily and more accurately in experiments. The observables chosen were the total multiplicity, average largest cluster size $a_{max}$ and a normalised variable $a_2$ which assume distinctly different values in liquid and the gas phases thus serving as order parameters of the transition. The  variation of these observaables with temperature is very much similar to that of entropy or excitation energy and hence the temperature derivatives behave as specific heat at constant volume ($C_v$) which is an established signature of phase transition. Transition temperature can be identified from the position of the maxima of these derivatives analogous to that of $C_v$. Multiplicity derivative ($dM/dT$) serves as a robust signal with very good performance even in presence of Coloumb interaction which is long range and thereby suppresses signatures of phase transition. This signature not only persists but gets enhanced after secondary decay of the primary hot fragments and thus can be easily detected in experiments which measures total multiplicity. This signal was proposed using the canonical thermodynamical model and was later confirmed by us using the lattice -gas model. This was very recently verified in other theoretical models as well as in experiment.  The other two observables concerning the derivatives of the largest and second largest  clusters also peak at the same temperature as that of $C_v$ and can be considered as signals of 1st order phase transition in one component system switching off the Coloumb interaction. It is sometimes easier to measure the size of largest cluster than to count the total multiplicity covering all the fragments produced in the experiment. The effect of source size and freeze-out volume on the transition temperature is also studied using the one-component model. The extension of this study for real nuclei will be a part of our future work.

\end{document}